\documentclass{article}

\usepackage{arxiv}

\usepackage[utf8]{inputenc} 
\usepackage[T1]{fontenc}    
\usepackage{hyperref}       
\usepackage{url}            
\usepackage{booktabs}       
\usepackage{amsfonts}       
\usepackage{nicefrac}       
\usepackage{microtype}      
\usepackage{lipsum}
\usepackage{amsmath}
\usepackage{graphicx} 

\title{Financial factors selection with knockoffs:\\fund replication, explanatory and  prediction networks}

\author{
  Damien Challet\\
  Université Paris-Saclay,  CentraleSupélec,\\
  Laboratoire de Mathématiques et Informatique pour la Complexité et les Systèmes, \\
  91192 Gif-sur-Yvette, France\\
  \texttt{damien.challet@centralesupelec.fr} \\
   \And
 Christian Bongiorno\\
  Université Paris-Saclay,  CentraleSupélec,\\
  Laboratoire de Mathématiques et Informatique pour la Complexité et les Systèmes, \\
  91192 Gif-sur-Yvette, France \\
   \And
 Guillaume Pelletier\\
Automated Market Making, BNP Paribas,\\ 20 boulevard des Italiens, \\75009 Paris, France\\

}

\begin{document}
\maketitle

\begin{abstract}
We apply the knockoff procedure to factor selection in finance. By building fake but realistic factors, this procedure makes it possible to control the fraction of false discovery in a given set of factors. To show its versatility, we apply it to fund replication and to the inference of explanatory and prediction networks.
\end{abstract}

\keywords{factors \and factor selection \and clustering \and lead-lag \and prediction \and backtest}

\subsection*{Dedication}

\textit{At a time when multi-processor desktop PCs were a rarity, there was a special breed of physicists that had mastered massively parallel computing techniques. Dietrich Stauffer was one of them. He also was a parallel researcher,  working simultaneously on many topics and papers.  This made a durable impression on the young researcher that I  (DC) was then.}

\textit{This submission required massive parallel computations (600 CPU cores for a few days). Dietrich would not have been impressed.}
  
\section{Introduction}
  
Factor hunting is a venerable task in Finance, either in an explanatory pursuit, or in a prediction setting. Factor may explain risk, performance and diversification. This endeavour has two distinct parts: finding candidate factors and selecting them. 

The celebrated Fama-French factors \cite{fama1993common,fama2015five} have two precious qualities: they are exactly a handful nowadays (5, up from 3 \cite{fama2015five}) and make sense from a financial point of view. Their small number is a sure way to avoid overfitting the data with too many factors, and their financial interpretation is straightforward in layman's terms. On the other hand, it is beyond doubt that they cannot possibly capture the subtleties of price return dynamics and that additional factors, possibly ephemeral ones, are needed.

Dropping the interpretability requirement, it is nowadays very easy to generate a large quantity of candidate factors from fundamental or alternative data. The difficult part is to select them in a statistically controlled way. Statistics suggests to regard factor selection among a given set of factors as multiple hypotheses testing: to each factor corresponds the null hypothesis that it is irrelevant. This opens the way to methods that are able to control the error rate when one selects factors \cite{benjamini1995controlling}. Trying to keep relevant factors begin unrealistic, one settles for a less ambitious aim. In finance in particular, one can tolerate a controlled fraction of wrong choices given the amount of noise contained in the data (whereas it is usual instead to accept an uncontrolled level of wrong findings unwittingly).

Whereas the usual methods of controlling the error rate are based on p-values, Ref.\ \cite{barber2015controlling} introduced the so-called knockoff procedure that dispense with p-values altogether. It consists in creating a fake but look-alike factor for each candidate factor, i.e., a knockoff (a copy of low quality) and to add all knockoffs to set of candidate factors.  Whereas one does not know if a candidate factor is relevant, the fake one is irrelevant by definition, which provides a way to estimate the fraction of wrong selection. Since its introduction, this method has spanned a flurry of new methods and has been steadily improved (see Refs.\  \cite{candes2018panning,gimenez2019improving,fan2020ipad,romano2020deep} for example). Here we apply them to raw financial asset price returns in three situations: fund replication, explanatory and prediction  networks.

\section{The selection problem\label{knockoff}}

Let us introduce some useful notations. Here, we are mostly interested in explaining some price return timeseries $r$ from a return matrix of $N$ assets $R$. In a linear regression setting, using the standard $y$ and $X$ vector and matrix notation, one writes

\begin{align}\label{eq:yXb}
    y  & =  X \beta + z \quad \text{with} \quad \begin{cases} 
    															y  \in \mathbb{R}^{T} \\
                                                                X  \in \mathbb{R}^{N\times T}         
    														\end{cases},
\end{align}
where $\beta$ and $z\in\mathbb{R}^N$ are respectively the factor the loadings and the residual vector. In this paper, the candidate factors only consist of price returns of a collection of assets.

A factor $i$ is selected whenever $\beta_i\ne 0$. It is highly likely that a least-squares optimization yields $\beta_i\ne 0$ for all values of $i$. A well-known way to restrict the number of selected factors  is to add an  $\mathbb{L}_{1}$ penalization to the least-square problem. The LASSO \cite{tibshirani1996regression} in particular can be written as 
\begin{align}
 \hat{\beta}(\lambda) = \underset{\beta  \in \mathbb{R}^{N}}{\text{argmin}}\left\lbrace \frac{1}{2} \|y-X\beta \|_{2}^{2}+\lambda \|\beta\|_{1} \right \rbrace,
\end{align} 
where the penalization constant $\lambda$ controls the number of selected components, i.e., of non-null elements of $\hat \beta$. 
Intuitively, relevant factors should be selected for larger values of $\lambda$ than irrelevant ones. The whole question is how to choose the constant $\lambda$ and according to which criterion.  Beyond the question of the number of selected factors, their relevance is also to be assessed, for example by assessing the fraction of wrong factors present a given selection.

Let us denote the set of index of the relevant factors, or equivalently, the selection set of the true factors by $S\subset \{1,\cdots,N\}$ and a given factor selection (obtained by whatever method) by $\hat S$ such that such  $i\in\hat S\Longleftrightarrow\hat\beta_i\ne 0$ and inversely.  
The False Discovery Proportion (FDP henceforth) of a given selection $\hat\beta$ is defined by:  

\begin{equation}
    \mathrm{FDP} = \frac{\#\{j: j\in \hat S\setminus S \}}{\#\{j:j \in \hat{S}\}}, 
\end{equation}
with the convention that $\mathrm{FDP}=0$ when no factor is selected. By definition, the False Discovery Rate is $\mathrm{FDR}=E(\mathrm{FDP})$. Then, a selection rule controls the FDR at level $q$ if $\mathrm{FDR} \leq q$. Note that this does not control the false rejection rate, i.e., the falsely rejected factors.

\subsection{Knockoffs}

Ref. \cite{barber2015controlling} proposes to solve the problem of factor selection by generating $N$ artificial factors that share some properties with the candidate factors, while being built to be irrelevant: they are called knockoffs of the original factors.   In practical terms, a factor and its knockoff should be as independent as possible  while the dependence between knockoff $i$ and $j$ mirrors the dependence between factor $i$ and $j$.

The quality of each factor is then assessed with some method, which essentially yields a statistics denoted by $Z$; we assume that a larger $Z$ is better. Let us denote by $\hat X $ the matrix of knockoff factors, and $Z_i$ and $\tilde {Z}_i$ the quality statistics of factor $i$ and of its knockoff.  Clearly, both distributions of $Z_i$  and $\hat{Z}_i$ are the same if $i$ is irrelevant, while one expects that $E(Z_i)>E(\hat{Z}_i)$ otherwise. 
Then the FDR is controlled according to the fraction of candidate factors such that $Z_i<\hat{Z}_i$ in a given selection set $\hat S$. For more details, the reader is referred to \cite{barber2015controlling,candes2018panning}.

For example, in the context of the LASSO problem, on average, relevant factors will be selected for larger values of $\lambda$ than their knockoffs, while there is no such ordering for irrelevant factors. Thus $Z_i$ can be the maximal value of $\lambda$  such that factor $i$ has $\beta_i\ne 0$. Other possibilities include the feature importance of factor $i$ in tree-based machine learning methods, which will be used later in this paper.

There are several ways to build knockoffs. Here, we use the so-called model-X method \cite{candes2018panning}, which can be used in the high dimensional case $N>T$. It rests on two assumptions: $y$ is independent of the irrelevant features and the distribution of
$[X,\tilde X]$ is invariant under the exchange of a factor and its knockoff.

Constructing knockoffs that respect these two conditions is generally hard. For the sake of computational speed, we use an approximate second order moment matching method from the {\tt knockoff  R} package \cite{knockoff_CRAN}. More powerful methods, based on deep learning \cite{liu2018auto,romano2020deep,sudarshan2020deep}, deal better with heavy tailed data and more complex dependencies. Given its additional computational burden and the already considerable computation power needed to produce our results, we leave it to a future study.

\section{Results}

\subsection{Fund replication}

Among some notable applications of this method to finance, fund replication is obtained by regressing the performance of the latter to the returns of strategies that it may potentially use, given the constraints of its prospectus (see e.g. \cite{weber2013hedge}). 

A straightforward application of knockoffs is to find out what sparse combination of assets can explain (and reproduce) the price returns of a given fund, $y_t$. We consider returns between adjusted daily close prices of 4400 US equities in the 2005-2016 period, with a calibration window of 252 days (about a year). In each calibration window, we remove assets which have any missing value. Even of the knockoff model-X method allows for having more assets than timesteps, $T=252$ and $N=4400$ would be quite ambitious. Thus, we perform 200 bootstraps (without replacement within a bootstrap) of 500 assets each.

\begin{figure}
\centering
	\includegraphics[width=0.5\textwidth]{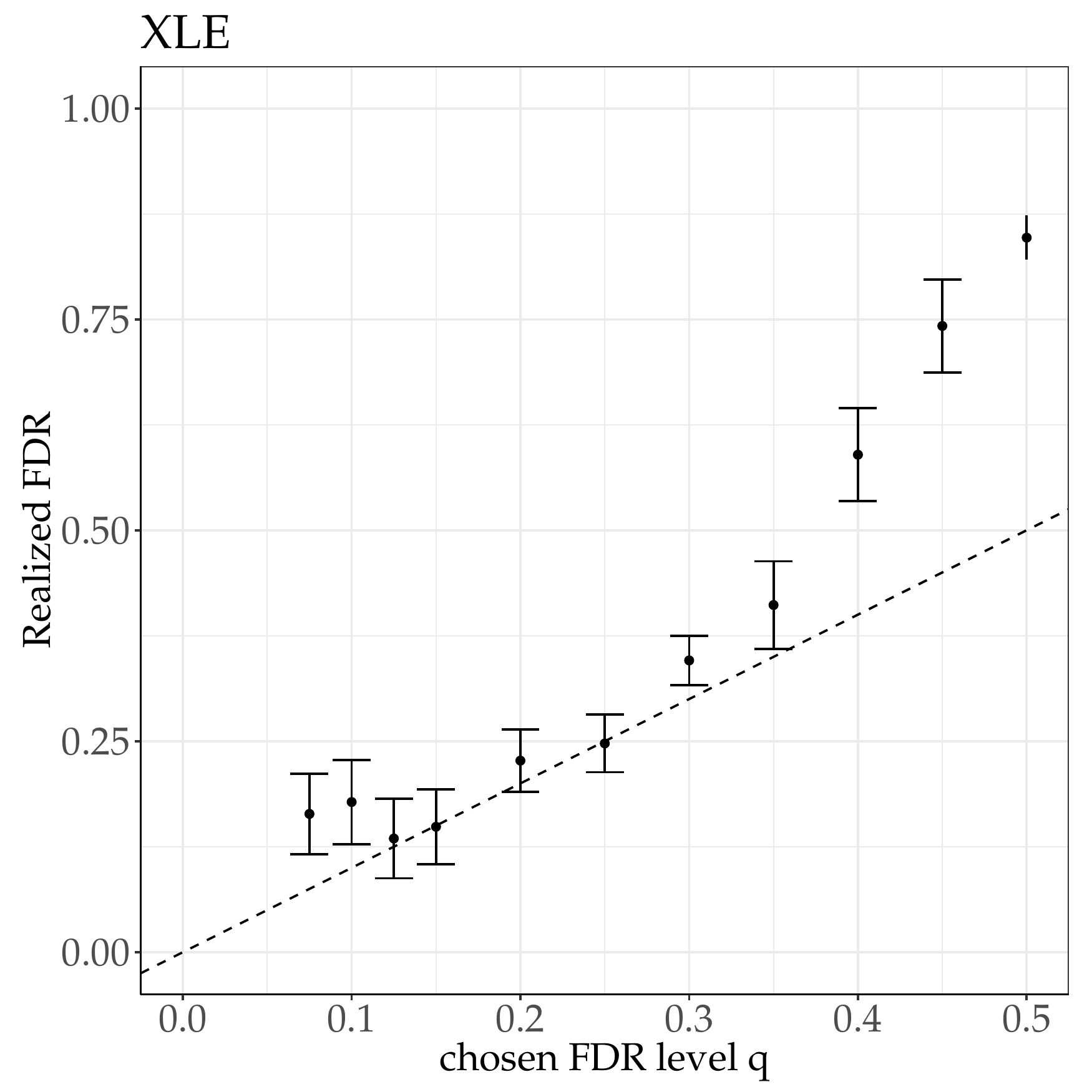}
\caption{Realized FDR versus chosen FDR when replicating the performance of XLE, an energy sector ETF. 200 bootstraps of 500 assets, 252 days of calibration length, average over 12 years.\label{fig:XLE_FDR}}
\end{figure}

As a simple illustration, we select an ETF which attempts to replicate the performance of the energy sector, XLE. For each calibration window, we compute the fraction between the number of selected assets that belong to the energy sector and the total number selected assets. In principle, this should give $1-\textrm{q}$, where $q$ is the chosen FDR level. We ran computations for various values of $q$ to check that knockoffs are consistent. Figure \ref{fig:XLE_FDR} reports that the realized FDR is close to $q$, except for large FDR (which would not be useful anyway), and for  $q<0.15$: this is because the typical number of assets belonging to the energy sector selected in a bootstrap reaches 1, which results in a boundary effect. In short, this simple test case confirms that knockoffs give meaningful results in a context in which the success rate is testable.

\subsection{Explanatory networks}

Knockoffs can be used to build explanatory networks of price returns, where there is a directed link from asset $j$ to asset $i$ ($i\ne j$) if $r_{j,t}$ contributes to explain $r_{i,t}$. In other words, the timeseries of returns of asset $i$ is the dependent variable $y$ and all the other assets are potential explanatory variables (respectively $y$ and $X$ in Eq.\ (\ref{eq:yXb}). Let us denote by $N_i^{(t_0,t_1)}$ the set of assets that explain the price returns of asset $i$ in a given calibration window $[t_0,t_1]$ at a given FDR level. In contrast to many other clustering methods based on symmetric measures (e.g. correlations) \cite{kullmann2000identification,giada2001data}, knockoff explanatory networks are directed: if asset $2$ and $3$ explain asset $1$, it may happens that asset $1$ does not belong to $N_2^{(t_0,t_1)}$. This implies that the knockoffs explanatory networks are more flexible than correlation-based methods and may contain more information. The downside is that more work is needed to extract clusters from these results, for example by performing network clustering, known as community detection in this stream of literature (see \cite{fortunato2010community} for a review). 

Here we focus on the time evolution of network properties especially between industries as defined by the GICS classification \cite{GICS}.
Because computing time scales as $N^2$ where $N$ is the number of assets to test, we focus on 200 US equities from 2000 to the end of 2017, which leads to reasonable computing time on several hundred CPU cores. We repeat the knockoff generation and selection process a given number of times ($100$ here) for each asset and for each calibration window and consider the union of all the selected factors. This is because the selection is empty for a sizable fraction of times, a known instability of knockoff generation (see e.g. \cite{gimenez2019improving}). We found that the knockoff selection using feature importance of candidate factors and knockoffs yielded by random forests gave the most factors at a given level of FDR.

\begin{figure}[tbh]
    \centering
    \includegraphics[width=0.4\textwidth]{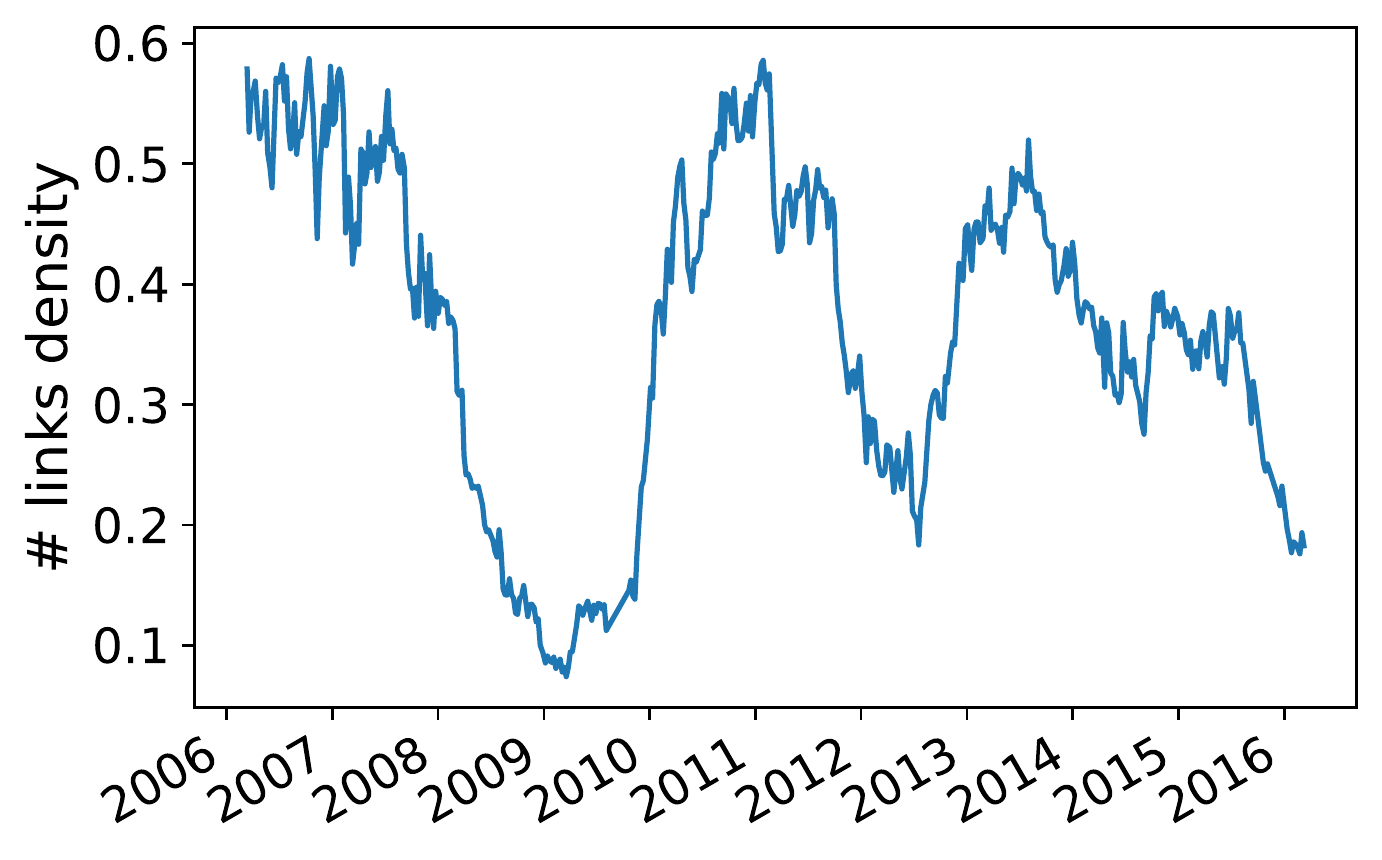}
    \includegraphics[width=0.4\textwidth]{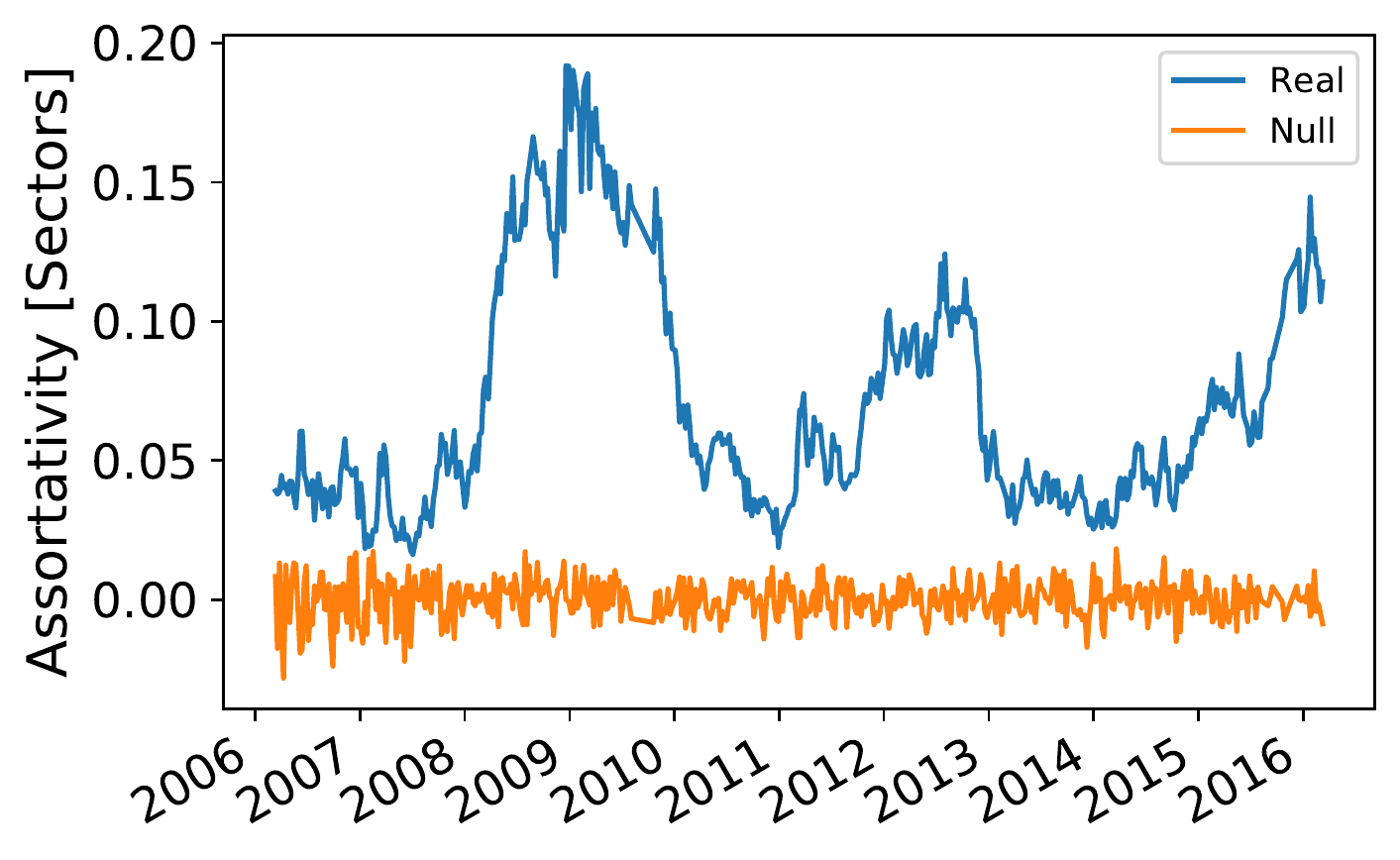}
    
    \includegraphics[width=0.4\textwidth]{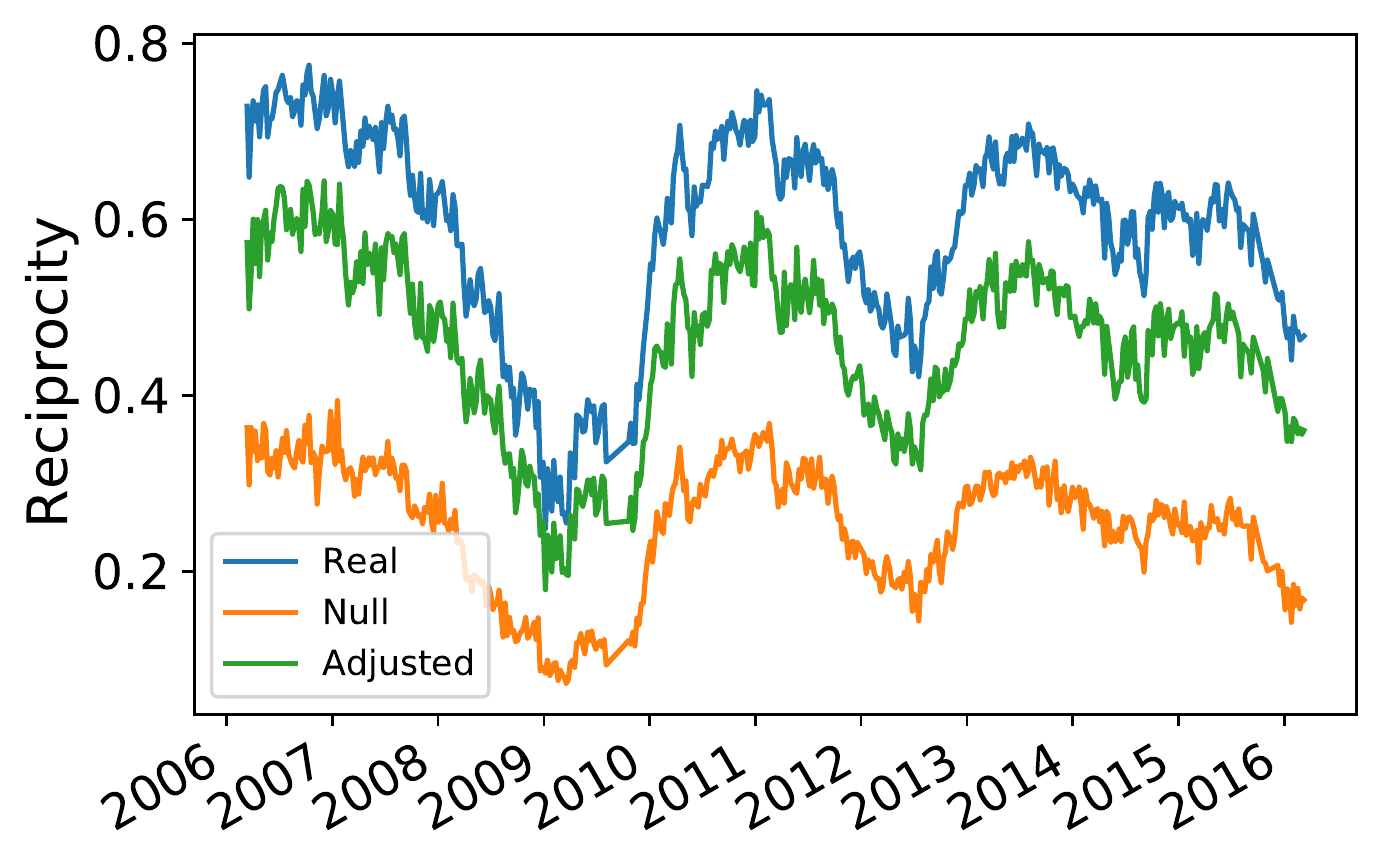}
    \includegraphics[width=0.4\textwidth]{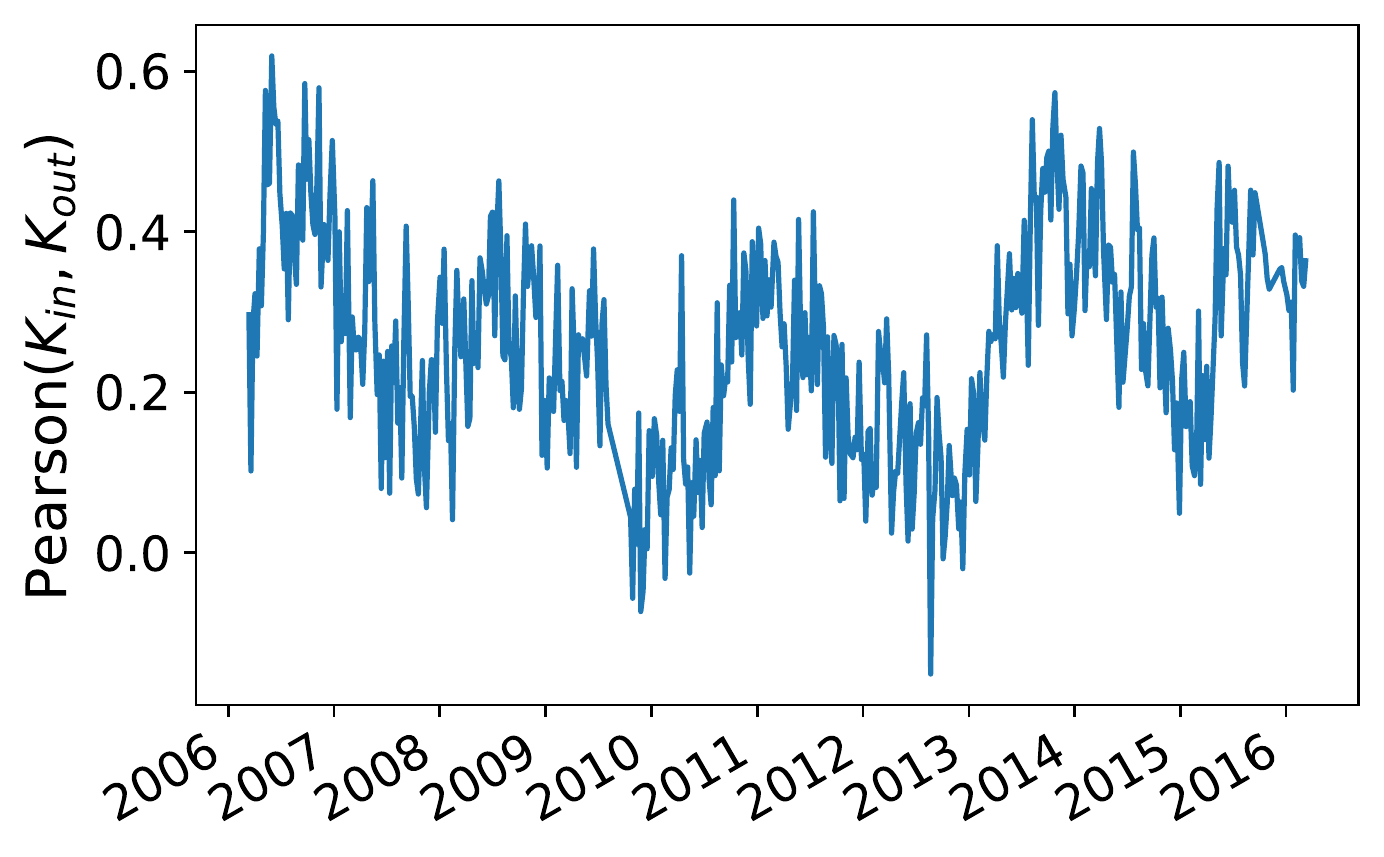}

    \caption{Upper left panel shows the link density; upper right panel shows the assortativity with sector classification, the figure reports the assortativity of a configuration model that preserves the in- and out-degree of the nodes; lower left panel show the link reciprocity, the figure report also the reciprocity of a configuration model that preserves in and out degree of the node and the adjusted estimator~\cite{garlaschelli2004fitness}; lower right panel show that Pearson coefficient between the in-degree and out-degree of the nodes. FDR$=0.1$, $T_{in}=300$  }
    \label{fig:netmetrics}
\end{figure}

Figure \ref{fig:netmetrics} reports the time evolution of four network synthetic measures. The link density, defined as the number of observed links divided by the all possible directed links ($N(N-1)$), is relatively high on average and has clear dips during the 2008-2009 and subsequent crises. To understand the origin of this phenomenon, we plot the reciprocity, i.e., the fraction of links that are bidirectional, which shows a similar behavior. Assortativity of links between sectors, which measures the propensity to establish links between nodes of the same sector (see \cite{newman2010networks}) is clearly significant, which is not surprising. It is noteworthy that it behaves in an opposite way from reciprocity \cite{newman2002email}. During times of crisis, links from the same industrial sectors are more likely to remain relevant.

\subsection{Prediction networks}

\begin{figure}
    \centerline{\includegraphics[width=0.4\textwidth]{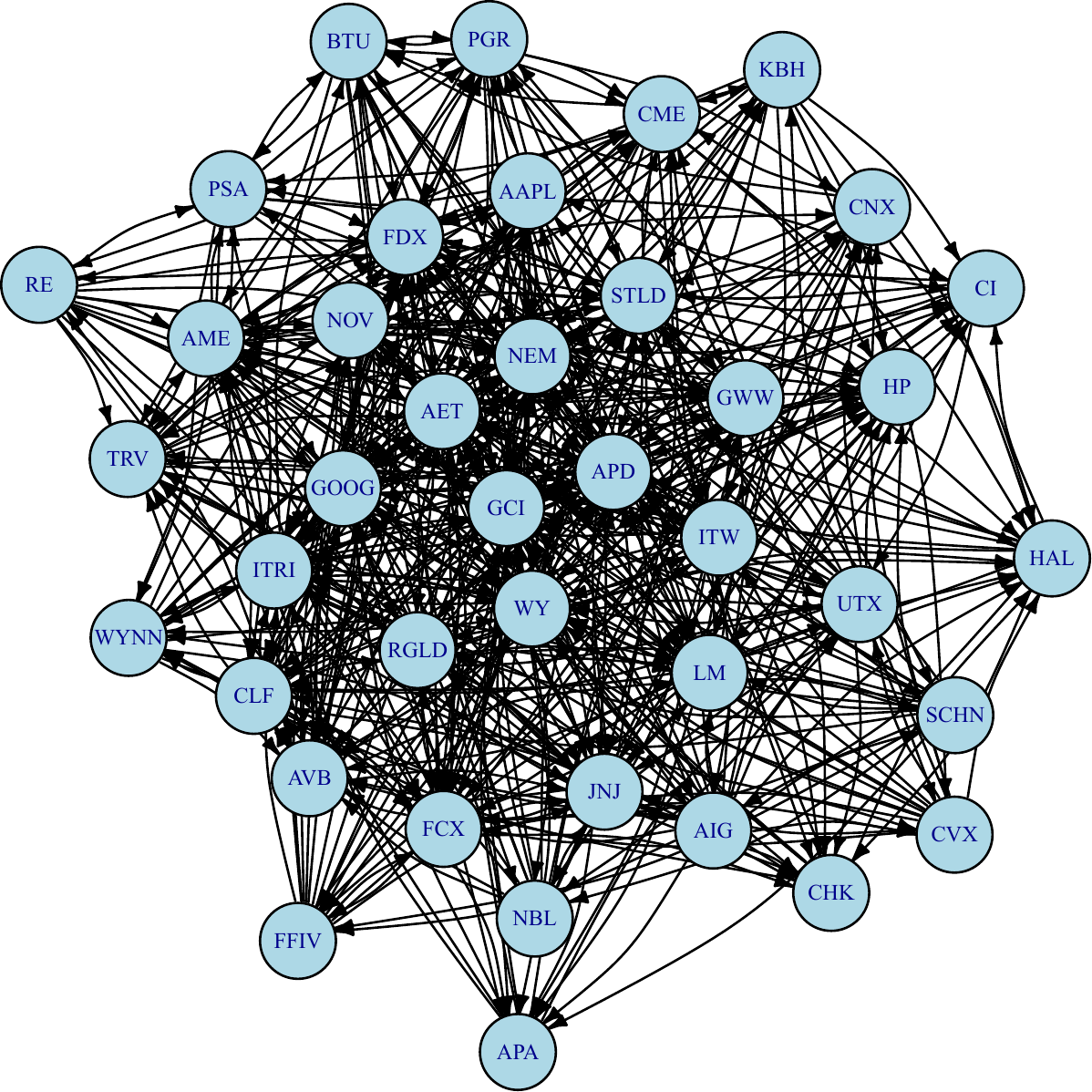}\ \ \ \ \includegraphics[width=0.4\textwidth]{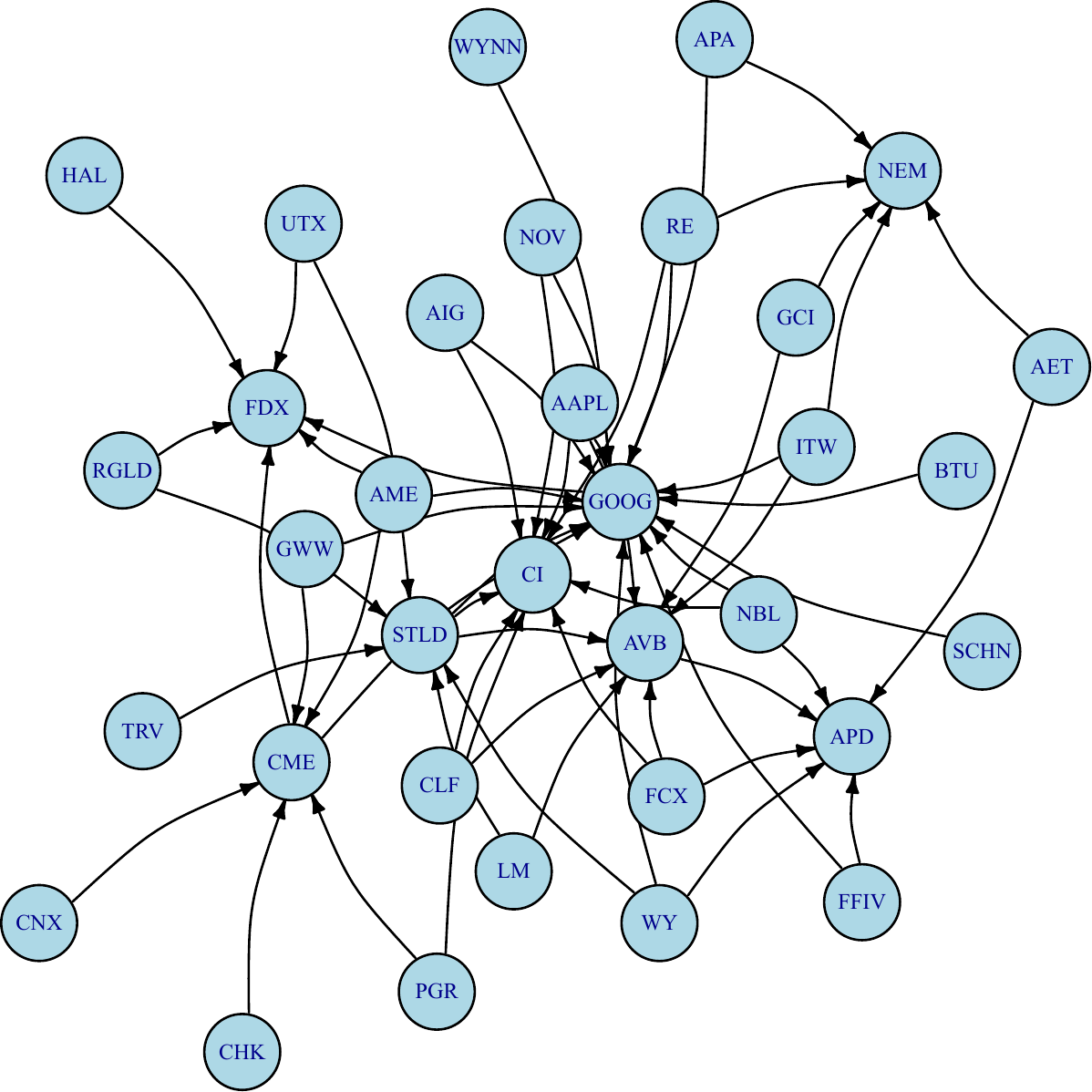}}

    \caption{Examples of prediction network inferred with knockoffs. $T_{in}=300$ trading days, period ending on 2006-03-13. Left plot: FDR$=0.3$ and 100 runs per asset; right plot: FDR$=0.2$ and 10 runs per asset. Selection via random forest variable importance.}
    \label{fig:pred_net}
\end{figure}

By lagging the predictors, one can explain future returns with previous ones. In other words, one builds prediction networks from knockoffs, as shown in Fig \ref{fig:pred_net}. Such networks are directed: for each asset  $i$ and time $t$, a collection of predictive assets $P_{i,t}$ links to $i$. We note that the in- out-links degrees is asymmetric: the number of out-going links is usually much smaller than the incoming links. What we call prediction networks here are lead-lag networks that are usually obtained by Granger causality (see e.g. \cite{papana2017financial} for a recent example among many) or by the asymmetry of lagged correlations  \cite{toth2006increasing,huth2014high,curme2015emergence}.

We use exactly the same setup as previously in the explanatory networks part.
Figure \ref{fig:pred_net} shows two examples of prediction networks obtained in the same calibration window, but with a different level FDR. The influence of that choice on the link density is very large: it is much more difficult to control an FDR set to 0.2 than 0.3 and hence much fewer links are found; almost no links are found for FDR$=0.1$. 

For time $t$, for each asset $i$ such that $P_{i,t}\ne \emptyset$, we apply a robust linear fit of $r_{i,t+1}$ with real returns of $P_{i,t}$ over the last $T_{in}$ days in-sample, and then predict the out-of-sample return for the next $\delta T=5$ days from the daily returns of each predictive assets. We first check the hit ratio of the  sign of the  prediction as a function of the number of elements in $P_{i,t}$, denoted by $k_{in}$. Quite remarkably, it decreases as a function of $k_{in}$ in a similar way for FDR$=0.2$ and $0.3$.

\begin{figure}
    \centerline{
    \includegraphics[height=0.4\textwidth]{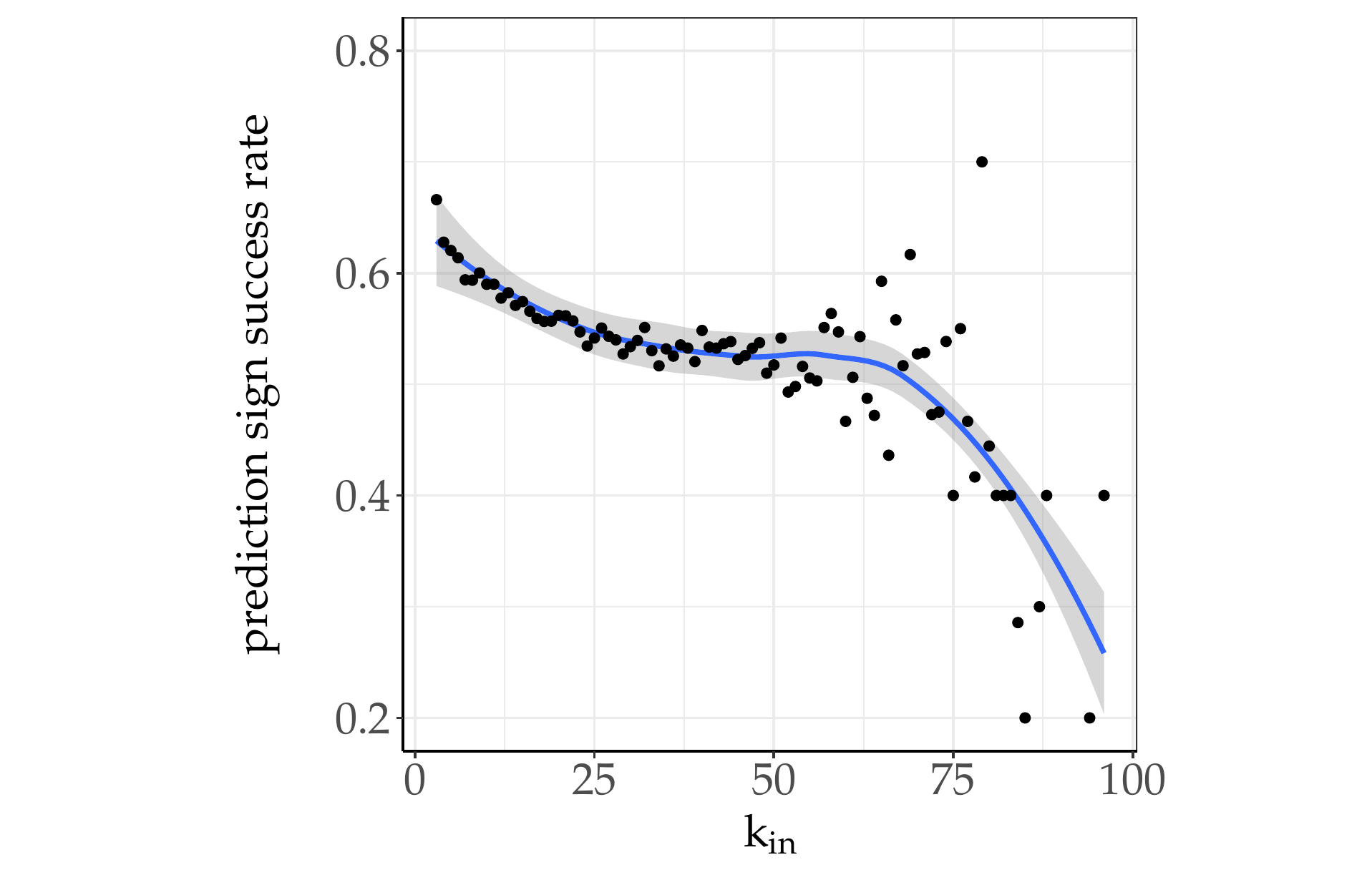}\hspace{-10ex}\includegraphics[height=0.4\textwidth]{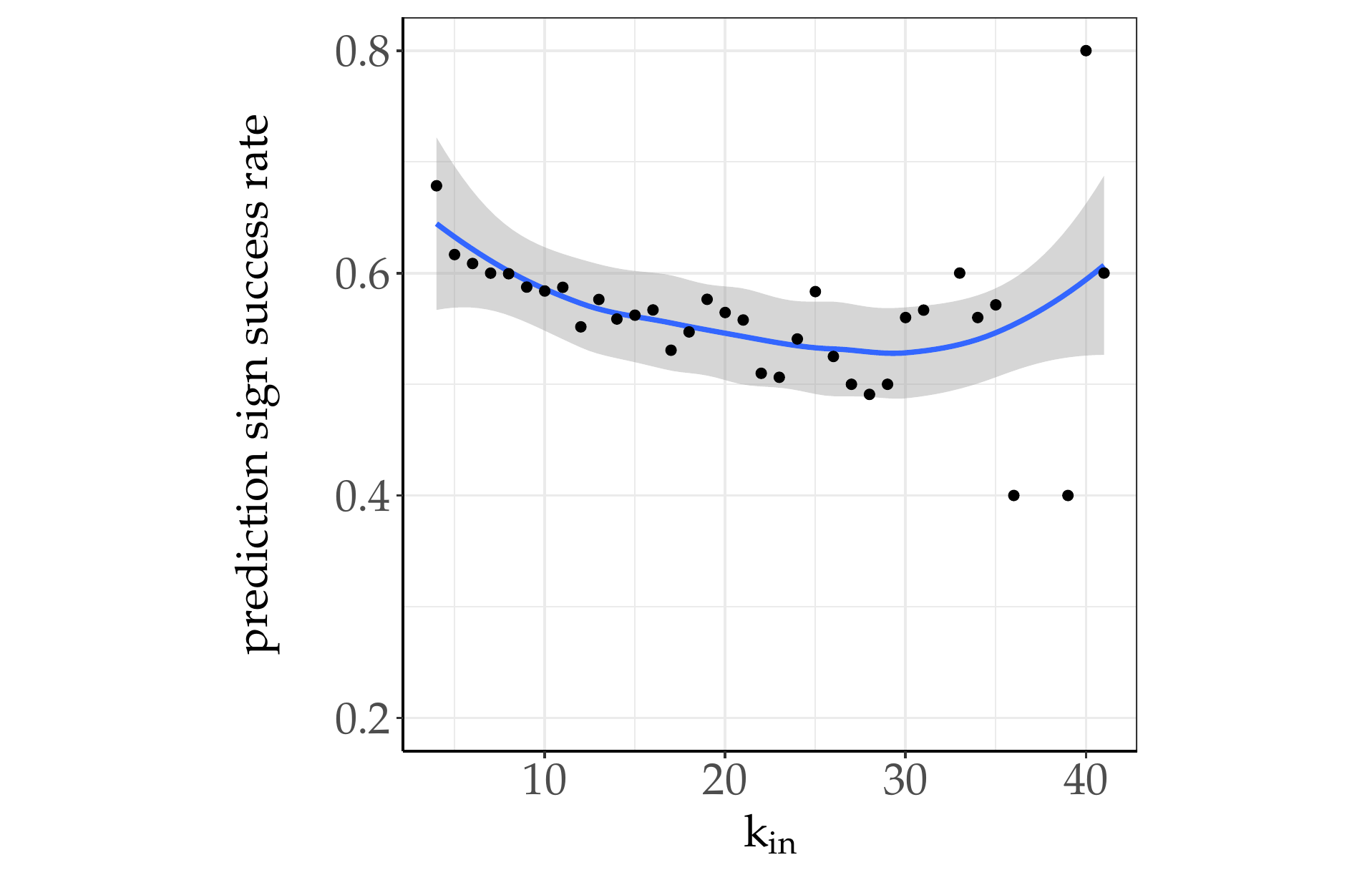}}
    \caption{Prediction hit ratio (out-of-sample) as a function of the number of assets $k_{in}$ that belong in predictors' set. Left plot: FDR=$0.3$; right plot: FDR$=0.2$. $T_{in}=300$; 200 assets. Blue curve: local average; gray area: standard error on the local average.}
    \label{fig:hitratio}
\end{figure}

While prediction networks can be used to build a trading strategy, the performance we report in the following cannot be considered as a proper back-test, as we use closing price both for computing returns and to open virtual positions, and because we do not include transaction costs. The point of this section is to show the information gain provided by filtering factors with knockoffs. We thus ran two experiments to assess the out-of-sample performance of the predicted assets: first we take the predicted assets, take a position according to the signs of their predicted returns, and compute the performance of equally-weighted portfolios, rebalanced at every timestep; in order to compare the information contained in the sign of the returns, we also add the performance of equally-weighed portfolios of long positions on predicted assets, which gives a performance similar to that of the market.  Both levels of FDR have acceptable performance before a saturation later on (2012-2017); this is in part because they provide long-short predictions, which worked better before 2015. We note however that FDR$=0.2$ leads to a better performance before 2015, as the signal provided by more trustworthy factors is better.

\begin{figure}
    \centering
    \includegraphics[width=0.5\textwidth]{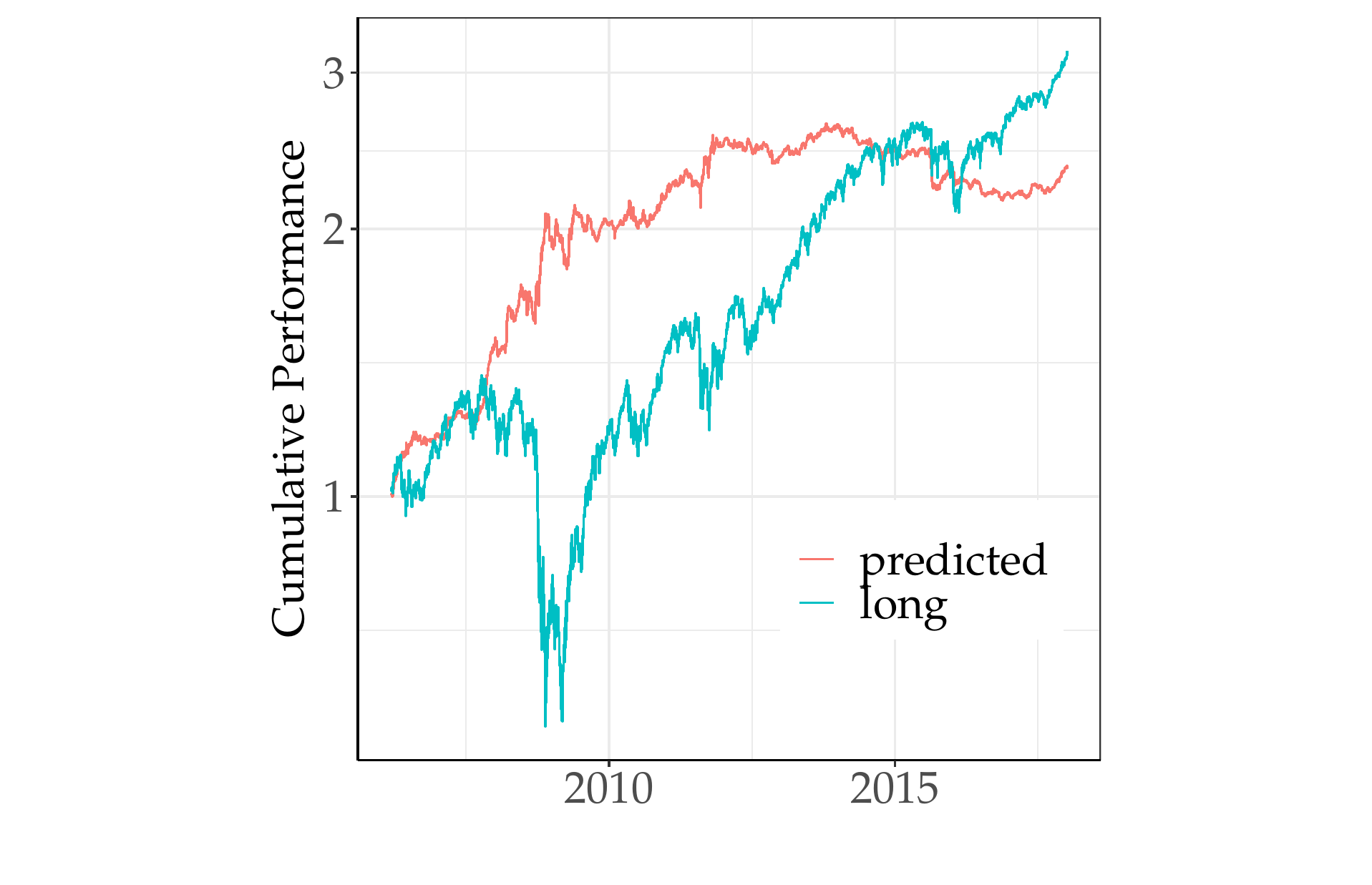}\hspace{-10ex}\includegraphics[width=0.5\textwidth]{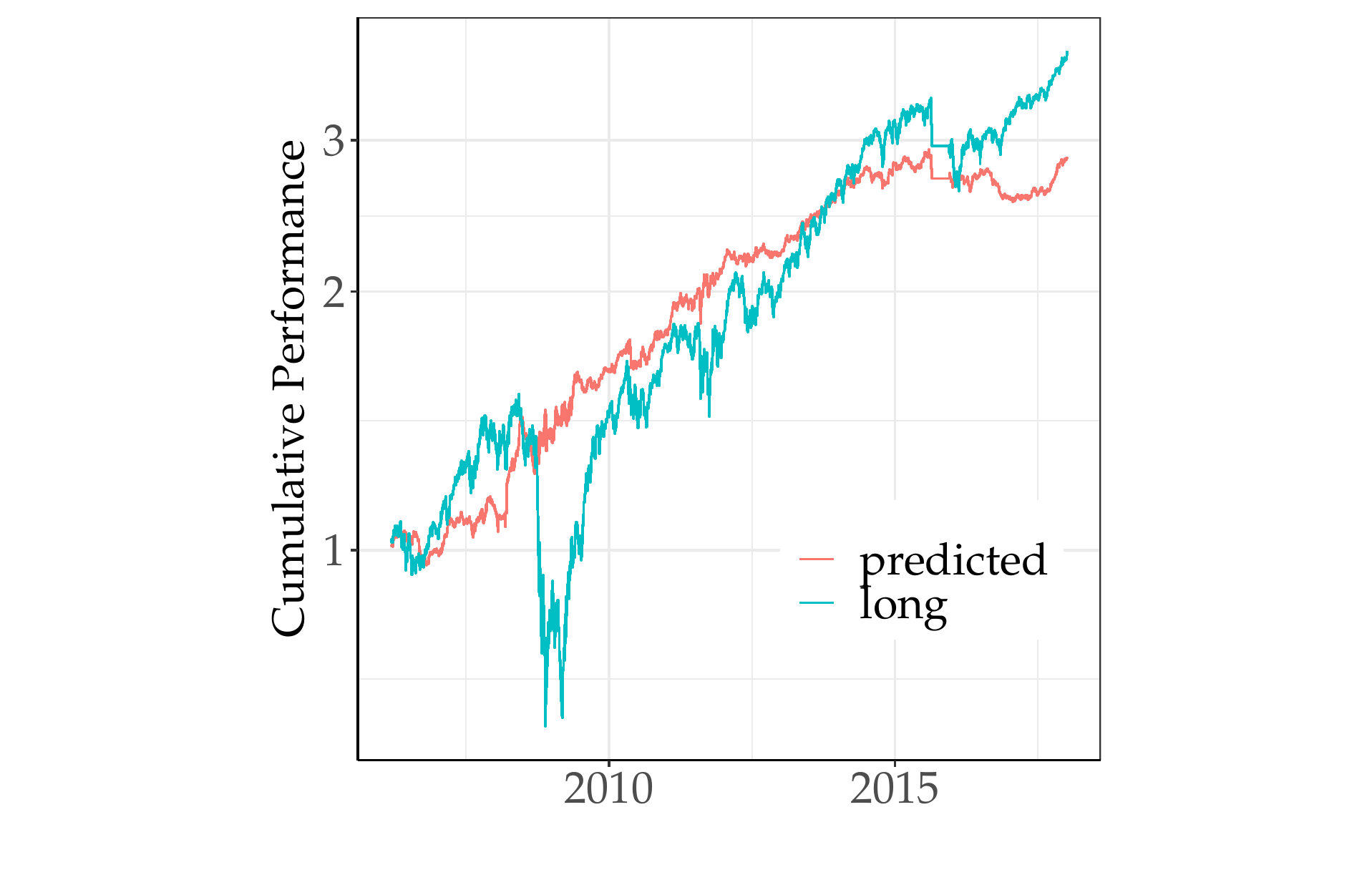}
    \caption{Cumulative performance from prediction networks computed for predicted assets with equally weighted portfolios either long-short (predicted), or long-only (long). $T_{in}=300$  Left plot: FDR$=0.3$; right plot FDR$=0.2$.}
    \label{fig:perf_Tin_raw}
\end{figure}

\begin{figure}
    \centering
    \includegraphics[width=0.4\textwidth]{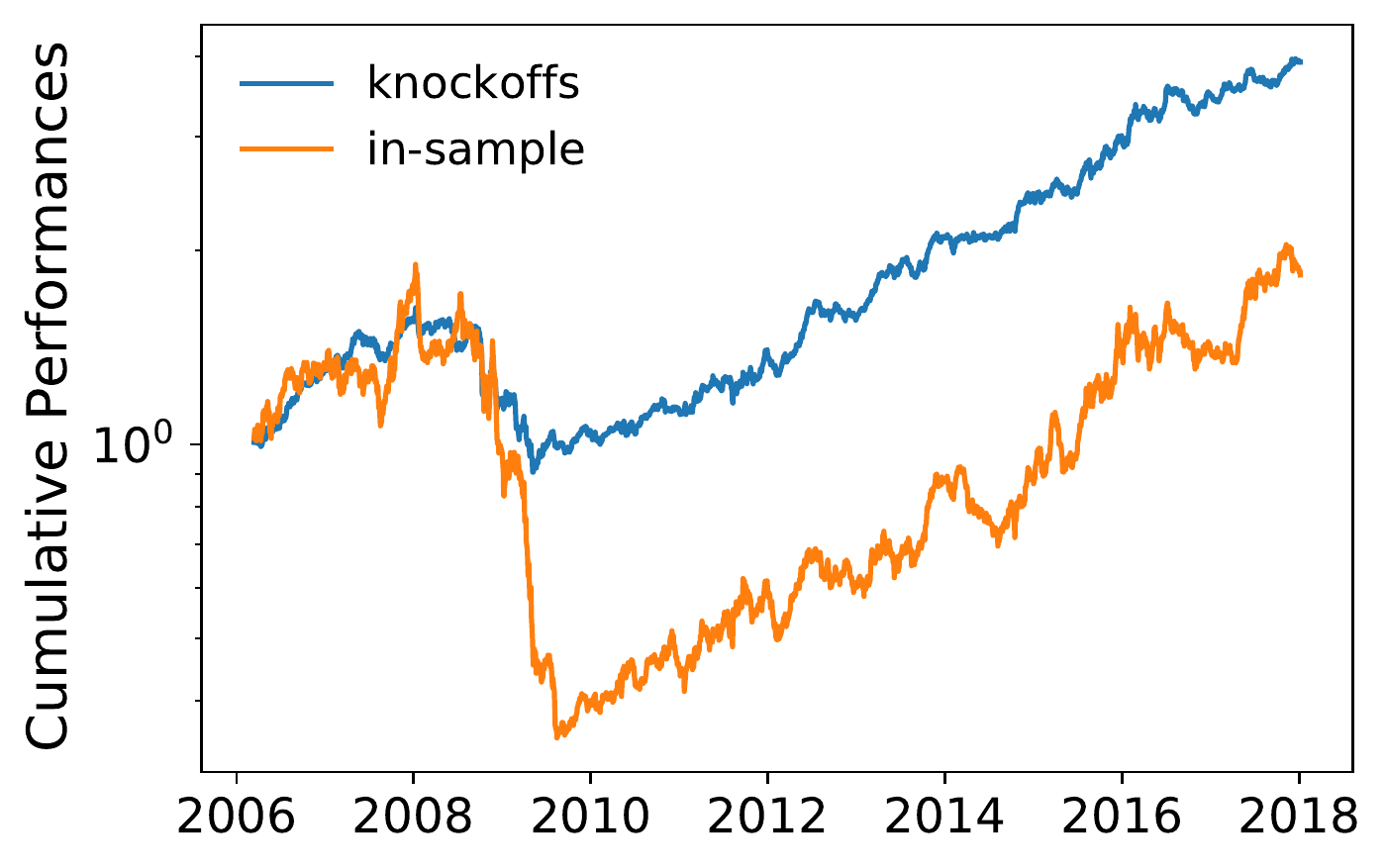}
    \includegraphics[width=0.4\textwidth]{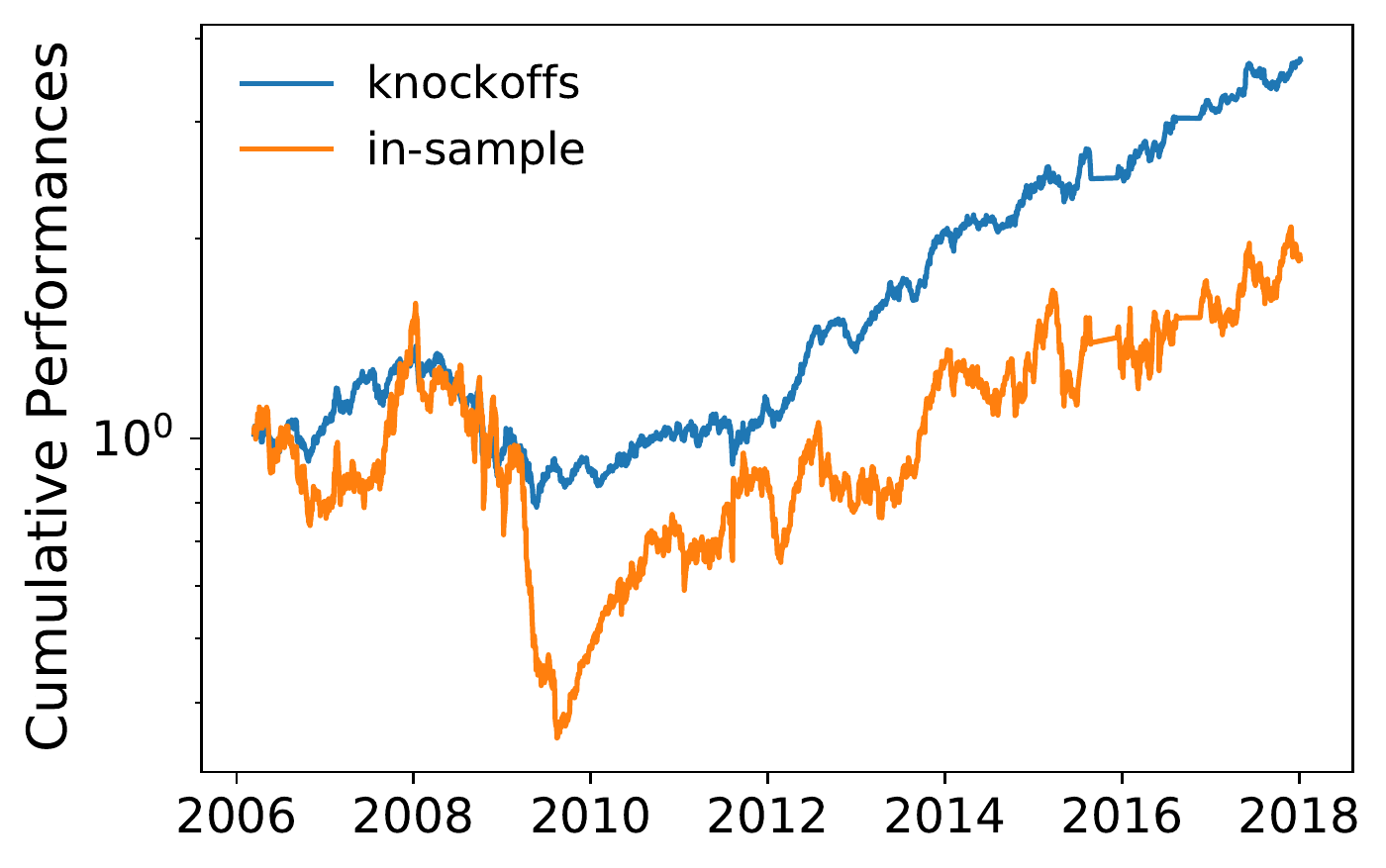}
    \caption{Cumulative performance from prediction networks  with mean-variance portfolios with target daily return equal to $0.005$, blue curve uses the knockoff prediction for the returns, the orange curve uses the average in-sample returns over the past 300 days. Left plot: FDR$=0.3$; right plot FDR$=0.2$.}
    \label{fig:perf_Tin}
\end{figure}

We then use the knockoffs selection and predictors in a more subtle way: for each day, we first compute the covariance matrix of all predicted assets over the last $T_{in}$, filtered with the BAHC method \cite{bongiorno2021covariance}. We then compute optimal long-short mean-variance portfolios at each time step with a fixed net leverage of 1. An interesting variation consists in using the predicted returns as the expected returns  instead of using the historical averaged returns. Figure \ref{fig:perf_Tin} shows the information gain provided by predicted returns. The difference of the gain profile with respect to equally weighted portfolios of Fig.\ \ref{fig:perf_Tin_raw} comes from the fact that fixing a net leverage to 1 induces a bias towards long positions.

\section{Conclusion}

Factor selection with knockoffs holds many promises in finance. This contribution only skims the surface by using price returns as factors. The originality of the knockoff method is that they  define directed networks where the presence of links is statistically controlled. Further work includes a better way to generate knockoffs, covariance matrix cleaning and applying knockoffs to other kinds of factors. 

\section*{Acknowledgements}

This publication stems from a partnership between CentraleSup\'elec and BNP Paribas.

This work was performed using HPC resources from the ``M\'esocentre'' computing
center of CentraleSup\'elec and \'Ecole Normale Sup\'erieure Paris-Saclay supported by CNRS and R\'egion \^Ile-de-France.



\end{document}